\documentclass[twocolumn,pra,aps,superscriptaddress]{revtex4-1}

\usepackage[latin9]{inputenc}
\usepackage{mathtools}
\usepackage{amsbsy}
\usepackage{amssymb}
\usepackage{amstext}
\usepackage{bm}
\usepackage{xcolor}
 \usepackage{soul}
\usepackage{ulem}
\definecolor{darkblue}{rgb}{0, 0, 0.5}
\newcommand{\RN}[1]{%
  \textup{\uppercase\expandafter{\romannumeral#1}}%
}
\usepackage[unicode=true,pdfusetitle,
bookmarks=false,
breaklinks=false,pdfborder={0 0 1},backref=false,colorlinks=true,allcolors=darkblue]
{hyperref}
\hypersetup{
	bookmarksnumbered=false,bookmarksopen=false}
\makeatletter

\@ifundefined{textcolor}{}
{%
	\definecolor{BLACK}{gray}{0}
	\definecolor{WHITE}{gray}{1}
	\definecolor{RED}{rgb}{1,0,0}
	\definecolor{GREEN}{rgb}{0,1,0}
	\definecolor{BLUE}{rgb}{0,0,1}
	\definecolor{CYAN}{cmyk}{1,0,0,0}
	\definecolor{MAGENTA}{cmyk}{0,1,0,0}
	\definecolor{YELLOW}{cmyk}{0,0,1,0}
}

\newcommand{\beq}{\begin{equation}}
\newcommand{\eeq}{\end{equation}}
\newcommand{\beqa}{\begin{eqnarray}}
\newcommand{\eeqa}{\end{eqnarray}}

\usepackage{amsmath}
\usepackage{graphicx}
\usepackage{amssymb}
\usepackage{txfonts,color}
\makeatother
\begin{document}
	
\title{Superoscillating quantum control induced by sequential selections}

\author{Yongcheng Ding}
\email{jonzen.ding@gmail.com} 	
\affiliation{Department of Physical Chemistry, University of the Basque Country UPV/EHU, Apartado 644, 48080 Bilbao, Spain}

\author{Yiming Pan}
\affiliation{School of Physical Science and Technology and Center for Transformative Science, ShanghaiTech University, Shanghai 200031, China}

\author{Xi Chen}
\email{xi.chen@csic.es} 	
\affiliation{Instituto de Ciencia de Materiales de Madrid, Consejo Superior de Investigaciones Científicas,
	Cantoblanco, 28049 Madrid, Spain}

\date{\today}

\begin{abstract}	
Superoscillation is a counterintuitive phenomenon for its mathematical feature of ``faster-than-Fourier", which has allowed novel optical imaging beyond the diffraction limit. In this article, we introduce a superoscillating quantum control protocol realized by sequential selections within the framework of weak measurement, which steers the apparatus (target) by repeatedly applying optimal pre- and post-selections to the system (controller).  Our protocol accelerates theadiabatic transport of trapped ions and adiabatic quantum search algorithm at a finite energy cost. We demonstrate the accuracy and robustness of the protocol in the presence of decoherence and fluctuating noise, and elucidate the trade-off between fidelity and rounds of selections. Our findings open new avenues for quantum state manipulation and wave-packet control using superoscillation across a range of quantum platforms.

\end{abstract}

\maketitle

\section{Introduction}
The concept of superoscillation was originally proposed as a footnote in a celebrated study on quantum measurement by Y. Aharonov et al.~\cite{AAV} in the late 1980s. The phenomenon occurs when a band-limited wave function varies arbitrarily faster than its fastest Fourier components, as allowed by its spectral content. In other words, a superoscillatory wave is a \textit{local} feature that oscillates at a much higher frequency than the overall frequency of the global band-limited wave. This counterintuitive but physically allowed property is dubbed as \textit{faster-than-Fourier} by M. Berry~\cite{FTF} and others~\cite{roadmap}, which offers promising optical applications by breaking the diffraction barrier~\cite{nrp}. To date, superoscillation has since been applied in super-resolution imaging and in manipulating nanoparticles, electrons, and atoms with spatiotemporally shaped light beams ~\cite{manipulation,space,Atom,time}. 

Meanwhile, with the advent of state-of-the-art quantum technologies, ingenious protocols based on weak measurements have become attainable for a variety of quantum applications~\cite{reviewweak}, including quantum steering~\cite{steering,gefen}, quantum tomography~\cite{tomo,tomosc}, geometric information~\cite{geometric,eraser,discord}, and transition detection~\cite{w2s,topological}. Moreover, sequential weak measurements can produce superoscillation by encoding the amplified weak value of an operator-arising from a repeatedly pre- and post-selected system- into the coupled quantum states of the apparatus~\cite{pointer}. This observation  inspires us to construct a general quantum control framework using superoscillation for the applications where quantum information encoding is essential for specific proposals of quantum state steering and wave-packet manipulation.

In this article, we propose a framework called superoscillating quantum control (SQC).  SQC manipulates the apparatus wave function efficiently by a measurement-induced superoscillating operator function, engineered through the optimal design of pre- and post-selections. We demonstrate two preliminary results: the nonadiabatic transport of a single trapped ion and the speed-up of a quantum search algorithm. At the same energy cost~\cite{energetic}, SQC delivers higher fidelity than conventional adiabatic control/computing by gently perturbing the ground state, resembling shortcuts-to-adiabaticity~\cite{stareview}, when the probabilistic cost is included. Furthermore, by investigating noise in open systems, we reveal the trade-off between the number of selection rounds and fidelity, identifying two types of mechanisms and their consequences for the speed and robustness of SQC protocols.

\section{Superoscillating quantum control}

As in  conventional applications of weak measurement to quantum control, quantum information is extracted from the system, and converted to classical information, typically with low signal-to-noise ratio. Thus, one can partially monitor the quantum state and make decisions based on measurement outcomes, resembling a form of closed-loop control.

On the contrary, SQC constitutes a novel framework wherein a quantum system is driven by sequential weak measurements. Two degrees of freedom, one severing as the system $\hat{S}$ and the other as the apparatus $\hat{A}$, are entangled by the most general Hamiltonian in the weak measurement regime, given by  $H_\text{SQC}=g\hat{S}\otimes\hat{A}$, which records a weak value $S_w=\langle f|\hat{S}|i\rangle/\langle f|i\rangle$ of the system operator on the apparatus, provided the pre-selected system $|i\rangle$ is successfully post-selected to $|f\rangle$ after the interaction. 

\begin{figure}
	\includegraphics[width=0.4\textwidth]{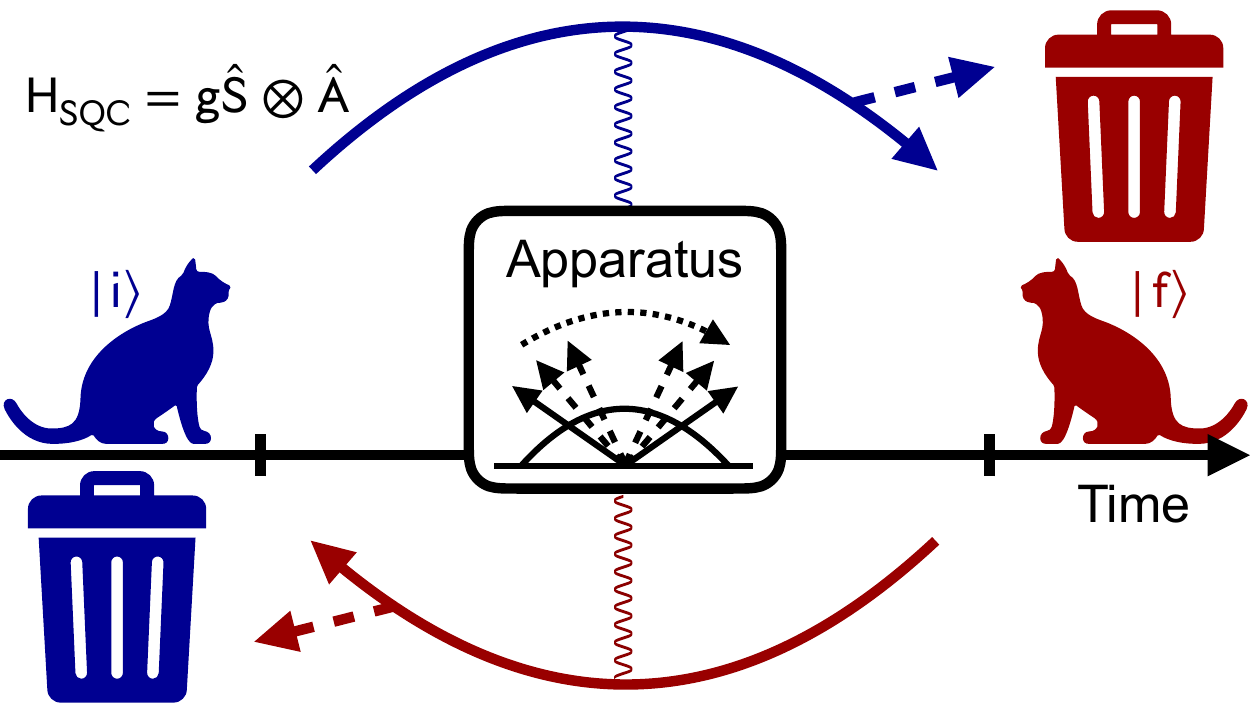}
	\caption{\label{fig:sqc} Schematic diagram of realizing SQC. The pre-selected system is coupled to the apparatus  via an interacting Hamiltonian, followed by post-selection to record a weak value on the apparatus. Through sequential selections, weak values accumulate to generate a  superoscillating function $f(\hat{A})$, which drives the  apparatus toward the desired target state. This enables efficient and robust quantum control of the apparatus wave function.}
\end{figure}

Apart from the original superoscillation with an ensemble of $N$ copies of the system and the apparatus, we perform sequential selections on the system between $|i\rangle$ and $|f\rangle$, as shown in Fig.~\ref{fig:sqc}, yielding a superoscillating operator function $f(\hat{A})$ on the apparatus wave function as
\begin{equation}
\label{eq:sqc}
|\Psi_A^F\rangle = |\langle f|i\rangle|^N\underbrace{\left[\cos\left(\frac{gT}{\hbar}\frac{\hat{A}}{N}\right)- i S_w \sin \left(\frac{gT}{\hbar}\frac{\hat{A}}{N} \right)\right]^N}_{=: f(\hat{A})} |\Psi\rangle,
\end{equation}
where $T$ is the total operation time uniformly divided by $N$ rounds of sequential selections. The weak value $S_w$ is chosen real and satisfies $S_w>1$ to ensure the mathematical properties necessary for superoscillation,  even though it can be complex in practice \cite{PRA76}. SQC occurs with the probability of $|\langle f|i\rangle|^{2N}$ since it discards the wave function once any selection fails. Although SQC is not costless, it offers us sufficient flexibility via the superoscillating function $f(\hat{A})$, enabling the design and acceleration of robust quantum control protocols through the accumulation of weak values.

\section{Transporting the Gaussian apparatus}

The  Gaussian wave function,  commonly used as an apparatus in quantum measurement scenarios, is coupled to a spin system via the Hamiltonian $H=g\hat{\sigma_x}\otimes\hat{p}$. However, such coupling often induces motional excitation of the apparatus. To mitigate this,   SQC can shift the apparatus without inducing final motional excitation. Using a minimal model, we exemplify SQC by the coherent transport of a single trapped ion.  In the Lamb-Dicke regime, where $\eta \sqrt{ \langle (a + a^\dagger)^2\rangle} \ll 1$ is satisfied, (as discussed Ref.~\cite{reviewtrappedion}),  the interaction between a two-level ion and a monochromatic photon mode of the light field can be described by the Hamiltonian,
\begin{equation}
\label{eq:ld}
H_{\text{LD}}=\frac{\hbar}{2}\Omega{\hat{\sigma}_+}\left[1+i\eta\left(\hat{a}e^{-i\nu t}+\hat{a}^\dagger e^{i\nu t}\right)\right]e^{i(\phi-\delta t)}+\text{H.c.},
\end{equation}
where $\Omega$ is the Rabi frequency, $\hat{\sigma}_+$ is the spin raising operator, and $\eta=kx_0$ is the Lamb-Dicke parameter with $x_0=\sqrt{\hbar/(2M\nu)}$ the characteristic motional length. Here, $\hat{a}$ and $\hat{a}^\dag$ are the annihilation and creation operators of the ion's motion, $\nu$ is the trap frequency, and  $\delta$ is the laser detuning. The parameters $\phi$,  $k$ and $M$ represent the laser phase, wave vector, and the ion mass, respectively.  The Lamb-Dicke Hamiltonian (\ref{eq:ld}) contains the red and blue sidebands. The first red sideband ($\delta=-\nu$)  reads
\begin{equation}
	H_{\text{rsb}}=\frac{\hbar}{2}\Omega\eta\left[\hat{a}\hat{\sigma}_+\exp\left(i\phi_r\right)+\hat{a}^\dag\hat{\sigma}_-\exp\left(-i\phi_r\right)\right],
\end{equation}
where $\phi_r$ is the laser phase. This interaction drives the transition $|n\rangle|g\rangle\leftrightarrow|n-1\rangle|e\rangle$ with effective Rabi frequency $\Omega\sqrt{n}\eta$. Similarly, the first blue sideband ($\delta=-\nu$) is:
\begin{equation}
	H_{\text{bsb}}=\frac{\hbar}{2}\Omega\eta\left[\hat{a}^\dag\hat{\sigma}_+\exp\left(i\phi_b\right)+\hat{a}\hat{\sigma}_-\exp\left(-i\phi_b\right)\right],
\end{equation}
inducing  another transition $|n\rangle|g\rangle\leftrightarrow|n+1\rangle|e\rangle$ with effective Rabi frequency $\Omega\sqrt{n+1}\eta$.
These interactions correspond to the well-known Jaynes-Cummings and anti-Jaynes-Cummings models, respectively. By employing a bichromatic laser field, combining the red and blue sidebands, with $\phi_r=-\pi/2$ and $\phi_r=\pi/2$, the Hamiltonian (\ref{eq:ld}) simplifies to
\begin{equation}
	\label{supep:bic}
	H_{\text{bic}}=H_{\text{rsb}}+H_{\text{bsb}}=-i\frac{\hbar\Omega}{2}\eta\hat{\sigma_x}\left(\hat{a}-\hat{a}^\dag\right).
\end{equation}
Substituting the expressions $\hat{a}=\sqrt{M\nu/2\hbar}(\hat{x}+i\hat{p}/M\nu)$ and $\hat{a}^\dag=\sqrt{M\nu/2\hbar}(\hat{x}-i\hat{p}/M\nu)$ into~\eqref{supep:bic}, we finally have $H=g\hat{\sigma_x}\otimes\hat{p}$ ~\cite{Lucas},  with coupling strength $g=\eta\Omega x_0$,  where $x_0=\sqrt{\hbar/2M\nu}$. In the weak coupling regime ($g=\eta\Omega x_0$), this enables high-fidelity, fast transport of a trapped ion using SQC.

\begin{figure}
	\includegraphics[width=0.45\textwidth]{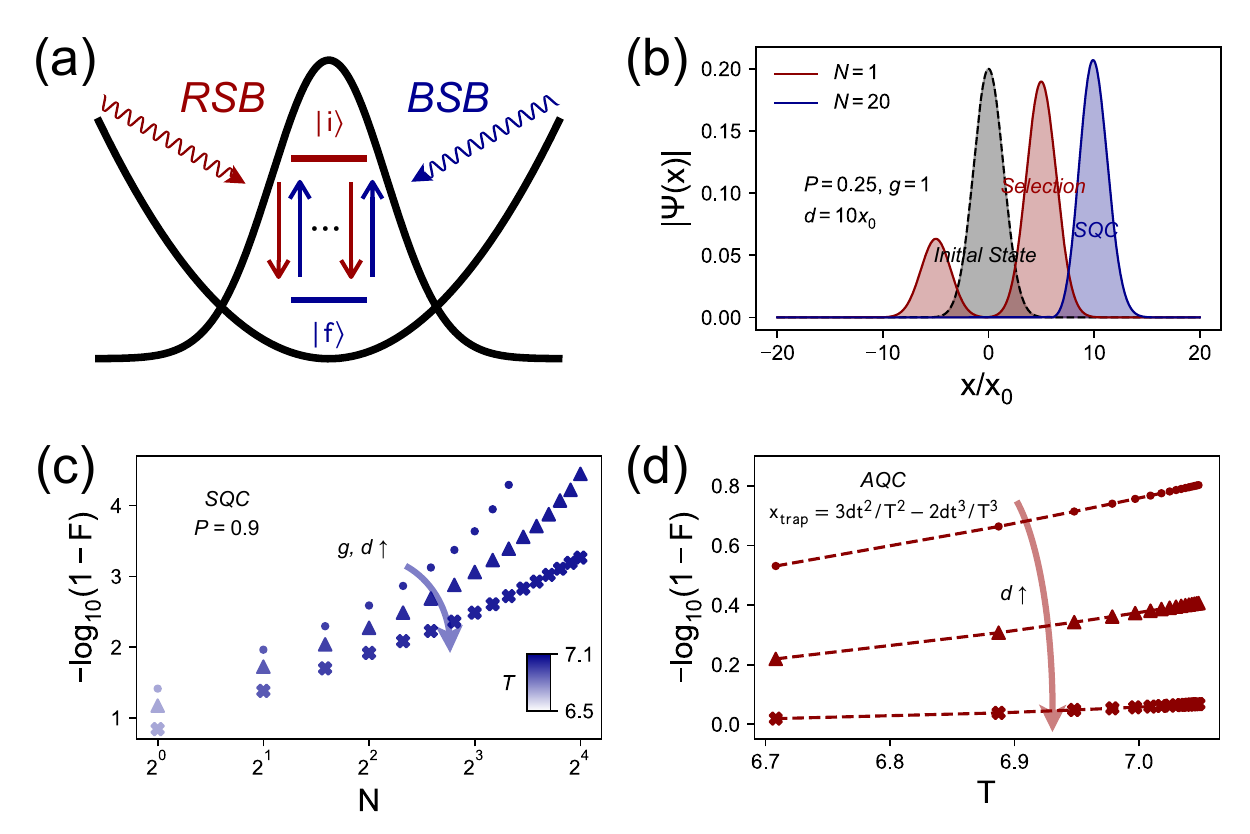}
	\caption{\label{fig:transport} (a) Schematic diagram of SQC in trapped ion quantum platform, where two lasers are tuned to the first red and blue sideband to couple the internal spin states to the motional mode. It realizes SQC with $N$ rounds of sequantial selections between $|i\rangle$ and $|f\rangle$ on the two-level system, transporting the ion to the target if all selections are successful. 
		(b) Comparison between single-shot strong measurement and SQC $(N=20)$ for transporting the ion by $d=10x_0$. Single-shot selection shows two non-overlapping red wave packets that cannot result in the unidirectional ion transport. The relevant dimensionless mass and trap frequency are $M=\nu=1$, and $P=0.25$. 
		(c) Fidelity dependence on $N$ via SQC for transport distances of $d=7.5x_0$ (dot), $10x_0$ (triangle), and $15x_0$ (cross) with $g=0.75,~1,~1.5$, and $P=0.9$. (d) Fidelity dependence on $T$ via AQC for the same transport distances as in (c). We move the center of the trap by defining $x_{\text{trap}}=3dt^2/T^2-2dt^3/T^3$.}
\end{figure}

Figure~\ref{fig:transport}(a) shows the setup: two lasers are tuned to the red and blue sidebands, coupling the internal and motional states. To realize SQC, we apply  $N$ rounds of sequential quantum-state pre- and post-selections by alternatively projecting the system into the initial state $|i\rangle$ and $|f\rangle$, and coupling the apparatus weakly to the system between two selections. According to Eq.~\eqref{eq:sqc}, the sequential selections on the system can exert a longstanding influence on the quantum state of apparatus, leading to the final state given by 
$$|\Psi^{F}_A\rangle=|\langle f|i\rangle|^Nf(\hat{p})|\Psi(x)\rangle=|\langle f|i\rangle|^N |\Psi(x+Ng\delta T\sigma_{xw})\rangle,$$
where $\sigma_{xw}=\langle f|\hat{\sigma_x}|i\rangle/\langle f|i\rangle$ is the weak value of $\hat{\sigma_x}$, and $\delta T=T/N$ is the interval between two selections. Each round shifts the apparatus by  $\delta x = gT\sigma_{xw}/N$,
accumulating to a net shift $\delta x = gT\sigma_{xw}$, without needing an ensemble of $N$ ions.

We are aware that SQC is not costless but only happens at the probability of $|\langle f|i\rangle|^{2N}$. Accordingly, it is necessary to analyze the relation between the probabilities and transport distance, and thus optimize the selections. For our concern, to obtain a real weak value  larger than $1$, we construct the two selected (initial and final) states as
\begin{eqnarray}
\label{eqn:if}
|i\rangle,~|f\rangle=\sqrt{\frac{1\mp\sqrt{1-p}}{2}}|0\rangle+\sqrt{\frac{1\pm\sqrt{1-p}}{2}}|1\rangle,
\end{eqnarray}
where $p$ is the probability of successful projection, giving optimal weak value $\sigma_{xw}=1/\sqrt{p}$. This results in the SQC protocol can shift the trapped ion by $d=gT/\sqrt{p}$, at cost $P=p^N$. At first glance, this seems problematic because the displacement is independent of $N$, and the total probability falls exponentially. That is, one can couple the internal states and the motional mode for the time interval of $T$ in a single shot of measurement. However, with strong coupling $(gT\gg x_0)$ ,  the post-selection no longer evolves as a translation operator on the motional wave function. From an intuitive perspective, the post-selection on the system after evolves the system-apparatus Hamiltonian determined by $\langle f|\exp(-igT\hat{\sigma_x}\otimes\hat{p})|\Psi(x)\rangle|i\rangle$, splitting the apparatus to the cat state (or kitty-state when two packets are not well separated):
\begin{eqnarray}
\label{eqn:w2s}
|\text{cat}\rangle = \frac{1+\sqrt{p}}{2\mathcal{N}}|\Psi(x-gT)\rangle+\frac{-1+\sqrt{p}}{2\mathcal{N}}|\Psi(x+gT)\rangle,
\end{eqnarray}
where $\mathcal{N}$ is the coefficient for normalization. Noteably, it presents the exact final quantum state of the apparatus, which is applicable for measurements with arbitrary strengths, ranging from weak to strong coupling. More detailed discussion can be found in Appendix \ref{Weakvaluecriteria}. The overlap between two wave packets is crucial to define the measurement transition, given by $\langle\Psi(x-gT)|\Psi(x+gT)\rangle=\exp(-\Gamma^2)$, where $\Gamma=gT/(\sqrt{2}x_0)$ is an interference factor. Using this, the expected net shift of the apparatus is calculated as 
\begin{equation}
\langle\delta x\rangle=\langle \text{cat}|x|\text{cat}\rangle=\frac{2\sqrt{p}gT}{1+p+\exp(-\Gamma^2)(-1+p)},
\end{equation}
which recovers  the weak-valued readout $\sigma_{xw} gT = gT/\sqrt{p}$ asymptotically as $\Gamma\rightarrow0$, and results in a bizarre readout $2\sqrt{p}gT/(1+p)$, that does not match the expectation value of $\hat{\sigma_x}$ in the initial or final state, in the opposite limit of strong measurement $\Gamma\rightarrow\infty$. More importantly, the net shift at weak measurement limit can be amplified when $p\rightarrow 1$. To preserve  unidirectional transport, we construct the SQC by periodically decoupling the system and the apparatus after every $\delta T=T/N$, keeping each measurement  in the weak regime. Fig.~\ref{fig:transport}(b)  compare  SQC with single-shot measurement for a target distance of $d=10x_0$ 
illustrating how SQC avoids motional splitting, where a low total probability of $P=0.25$ is chosen to increase result contrast, 

In Fig.~\ref{fig:transport}(c), we set the probability to $P=0.9$ and vary the number of selections from $N=1$ to $16$ to test the ideal performance of our protocol under different measurement strengths. We fix the operation time corresponding to $N$ in each set of settings by letting the coupling strength be proportional to the transport distance as (\romannumeral1): $g=0.75,~d=7.5x_0$,  (\romannumeral2): $g=1,~d=10x_0$,  (\romannumeral3): $g=1.5,~d=15x_0$. The results support our theory that the fidelity increases with $N$ and decrease with transport distance $d$.

 We notice that the extra time cost (compared with $T=6.708$ for $N=1$) for SQC is not remarkable since the nonoverlapping wave packet on the left is almost negligible with near-parallel selection ($P\approx 1$). Furthermore, we highlight that the SQC converges to the lossless expectation value amplification (also the eigenvalue $\langle\hat{\sigma_x}\rangle=1$) with near parallel selection of the system states. In this case, the cat state of apparatus~\eqref{eqn:w2s} reduces to a single wave packet $|\Psi(x-gT)\rangle$, and equalizes the weak value $1/\sqrt{p}$ or the the previous bizarre readout $2\sqrt{p}/(1+p)$ in either limit.  Importantly, SQC provides an alternative shortcut to adiabatic transport, distinct from inverse engineering and compensating force methods \cite{Erik,An,transportreview}.
 Finally, Fig.~\ref{fig:transport}(d) benchmarks SQC against the adiabatic quantum control (AQC) , where the center of  harmonic trap follows the trajectory $x_{\text{trap}}=3dt^2/T^2-2dt^3/T^3$ \cite{Erik}.  
Our protocol significantly outperforms such AQC in fidelity $F=|\langle\Psi(x,T)|\Psi_G\rangle|^2$, where $|\Psi_G\rangle$ is a Gaussian ground state at the target site, and prevails even when taking the probability of the selections into account.

\section{Rotating the qubit apparatus}

As we emphasized, SQC provides a general framework for quantum control that can be applied to arbitrary apparatus. In particular, it enables the accumulation of weak values on the angular parameters of a qubit wave function, opening avenues for quantum information processing. Interestingly, SQC can emulate Grover's search algorithm~\cite{haystack}, which locates a target state $|t\rangle$ from a database $|\Psi\rangle$ with a complexity of $\mathcal{O}(\sqrt{N_G})$ in digitized quantum computing, where $N_G$ is the number of entries.

Consider the interaction Hamiltonian  $H=g\hat{\sigma}\otimes\hat{J}$ , which couples the system and the apparatus. Through sequential post-selections, this induces the SQC effect on the apparatus by accumulating weak values $\sigma_w=\theta/g\delta T$.  The final apparatus state becomes
\begin{eqnarray}
\label{eqn:sog}
|\Psi_A^F \rangle=  |\langle f|i\rangle|^N f(\hat{J})|\Psi(\theta/2)\rangle= |\langle f|i\rangle|N |\Psi(\theta/2+N\sigma_wg\delta T)\rangle,
\end{eqnarray}
where $f(\hat{J})$ acts analogously to the Grover operator $\hat{G}^N=(U_\Psi U_t)^N$.  The oracle operator $U_t=I-2|t\rangle\langle t|$ and the diffusion operator $U_\Psi=2|\Psi\rangle\langle\Psi|-I$ flip the phase of the wave function in the target subspace and in the initial database, respectively. 

\begin{figure}
	\includegraphics[width=0.47\textwidth,height=3.1cm]{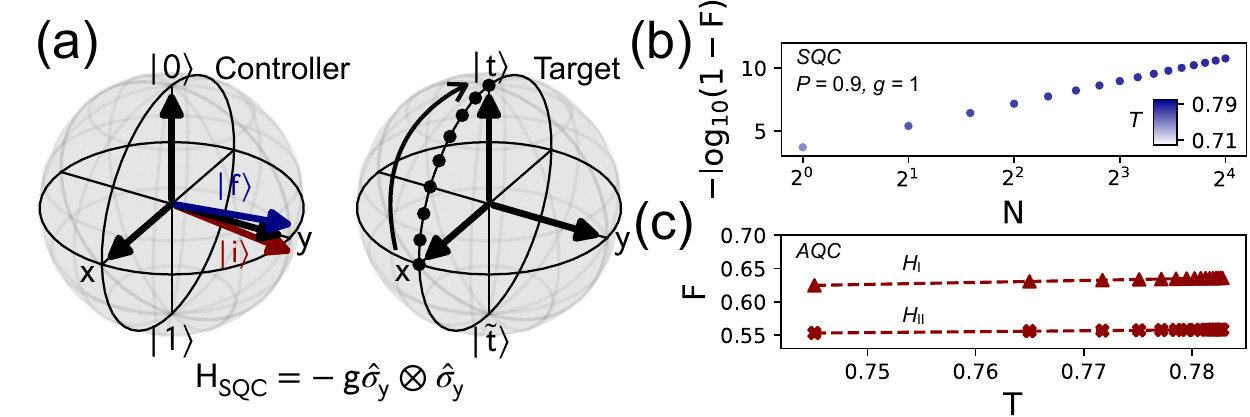}
	\caption{\label{fig:grover} (a) Schematic diagram of quantum search algorithm using SQC~\eqref{eqn:sog}. $N=8$ rounds of selections on the controller qubit, driven by the spin-spin interaction Hamiltonian $H_{\text{SQC}}=-g\hat{\sigma_y}\otimes\hat{\sigma_y}$, rotate the target qubit from the initial database $|\Psi\rangle=|+\rangle$ to the target $|t\rangle=|0\rangle$.  (b) Fidelity dependence on $N$ via SQC. The probability of SQC and coupling strength are set to $P=0.9$ and $g=1$, respectively. (c) Fidelity dependence on $T$ via AQC. The energetic cost is bounded, and the AQC Hamiltonians, see Eqs.~\eqref{eq:type1} (red triangle) and \eqref{eq:type2} (red cross), are evolved for $T$ that depends on the rounds of selections in SQC algorithm.
	}
\end{figure}

More intriguingly, SQC resembles the adiabatic version of Grover's algorithm for perfect query of $|t\rangle$.
In Fig.~\ref{fig:grover}(a), we demonstrate the algorithm by using a two-qubit system, one for the controller qubit (system) and one for the target qubit (apparatus), as a minimal quantum model that searches for the target state $|t\rangle=|0\rangle$ out of $N_G=2$ entries. Assuming that the initial database is $|\Psi\rangle=|+\rangle=\sin(\pi/4)|0\rangle+\cos(\pi/4)|1\rangle$, we can achieve a perfect query by setting $g\delta T\sigma_{yw}=\pi/4N$, which is unattainable in the standard Grover framework. Accordingly, we use the spin-spin interaction for implementing the SQC, which is the Heisenberg type $H_{\text{SQC}}=-g\hat{\sigma_y}\otimes\hat{\sigma_y}$~\cite{review2021}. We add the relative phase term $\exp(-i\pi/2)$ to the coefficients of Eq.~\eqref{eqn:if}, and a trade-off remains between the selection probability and weak value accumulation.

To quantitatively compare the energy cost of SQC and AQC, we consider time-dependent Hamiltonian $H(t)=[1-\lambda(t)]H_i+\lambda(t)H_f$, with $\lambda(t)$ ranging from 0 to 1  over total time $T$. We benchmark two types of AQC Hamiltonians.  Type-\RN{1} for the specific quantum search~\cite{stagrover} is given by
\begin{eqnarray}
\label{eq:type1}
H_{\RN{1}}(t)&=&(1-t/T)\Omega\hat{\sigma_x}+(t/T)\Delta\hat{\sigma_z},
\end{eqnarray}
with Rabi frequency $\Omega$, and the detuning $\Delta$. A more general type-\RN{2} Hamiltonian~\cite{aqcgrover} is 
\begin{eqnarray}
\label{eq:type2}
H_{\RN{2}}(t)&=&(1-t/T)K(I-|\Psi\rangle\langle\Psi|)+(t/T)K(1-|t\rangle\langle t|),
\end{eqnarray} 
where $K$ is a scaling factor. To evaluate performance, we track the fidelity 
 $F=|\langle\Psi(T)|t\rangle|^2$ and fix the success probability of SQC at  $P=0.9$, using coupling strength $g=1$. Fig.~\ref{fig:grover}(b) shows the fidelity 
 versus  the different selection number $N$. It is evident that the fidelity can be improved without increasing the operation time $T$ by using a larger Rabi frequency and detuning in the type-\RN{1} Hamiltonian, or equivalently scaling up $K$ in the type-\RN{2} Hamiltonian. Therefore, it is critical to define the energey cost~\cite{energetic} that bounds the input energy to the system for a fair comparison. We use the Frobenius norm of the total Hamiltonian to define the instantaneous cost of the evolution $\partial_t C=||H(t)||$, integrate and average it over the operation time as $C=\frac{1}{T}\int_0^T||H(t)||dt$. This yields $C_{\text{SQC}}=2g$ for the superoscillating quantum search. Assuming $\Omega=\Delta$, we derive the costs of the adiabatic algorithms, i.e., type-\RN{1} and type-\RN{2} Hamiltonian, as $C_{\RN{1}}=\Omega[2\sqrt{2}-\log(-1+\sqrt{2})+\log(1+\sqrt{2})]/4$ and $C_{\RN{2}}=K[4+3(-\log1+\log3)]/8$. 
 Fig.~\ref{fig:grover}(c) shows the energy cost of adiabatic algorithms compared to SQC, along with the fidelities of both Hamiltonians for the adiabatic quantum search within time $T$, corresponding to $N$ selection rounds.
 Numerical results demonstrate that SQC dramatically accelerates the adiabatic Grover's algorithm, even when both probabilistic and energetic costs are considered. We analyze its extension to arbitrary databases using multi-qubit system in the Appendix~\ref{MSgate}. The model is physically feasible in the trapped ion platform as well, where M\o lmer-S\o rensen gates serve naturally as analog simulators for the interaction.

\section{Discussions}

\subsection{Possible experimental implementation}

Both examples in ideal simulations point to the conclusion that SQC performs better with larger $N$. For implementation in the laboratory, we must model errors to evaluate its robustness and verify whether the previous result still holds under non-ideal conditions. First of all, systematic errors in the coupling coefficient and selection errors on the system-i.e., deviations from the ideal pre/post-selection-are mitigated by a factor of $1/\sqrt{N}$. This result, proven from an ensemble perspective, shows a significant advantage of SQC without challenging the earlier conclusion.

\begin{figure}
	\includegraphics[width=0.48\textwidth,height=3.2cm]{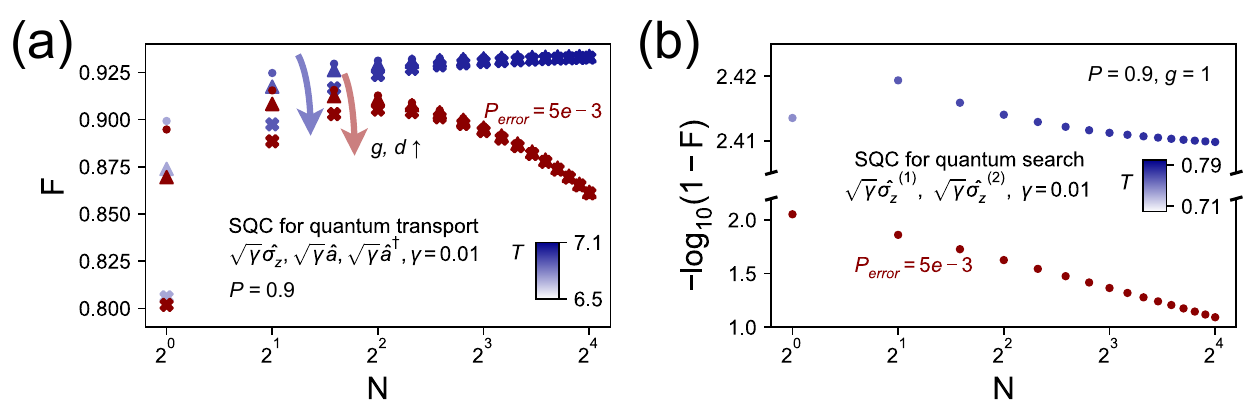}
	\caption{\label{fig:error} (a) Fidelity  vs. $N$ in SQC-based quantum transport, with quantum noises characterized by dephasing and damping rates of $\gamma=0.01$. Simulation results for the open quantum system are multiplied by $1-P_{\text{error}}=0.995$ after each selection (red markers) to mimic atomic loss induced by imperfect projections. Parameters and markers are the same as in Fig.~\ref{fig:transport}(c). (b) Fidelity vs. $N$ in SQC-based quantum search, with local dephasing modeled by  $C_{\text{TLS}}=\sqrt{\gamma}\hat{\sigma_z}$ on the controller and target qubits. Results are shown for perfect (blue) and imperfect (red) projections with the same settings as in (a) and in Fig.~\ref{fig:grover}(b).}
\end{figure}

Next, we further evaluate SQC in open quantum systems, accounting for imperfect selections that can introduce atomic loss and quantum noises, thus reducing fidelity by a fixed proportion $P_{\text{error}}$. These mechanisms can create a trade-off between the fidelity and the number of selections: fidelity first increases with $N$, reaches a maximum, and then decreases beyond a critical value. Notably, the trade-off due to quantum noise does not necessarily appear in all settings.  As $N$ increases, SQC retains its advantage: the fidelity increases with added operation time, but this is counteracted by the accumulation of decoherence. Whether the trade-off appears depends on which factor dominates: fidelity gain from SQC or fidelity loss due to noise.

We model the quantum noises by collapse operators $C_{\text{TLS}}=\sqrt{\gamma}\hat{\sigma_z}$ and  $C_\text{HO}=\sqrt{\gamma}\hat{a}~,\sqrt{\gamma}\hat{a}^\dag$ in the ion transport problem to characterize dephasing of internal state and damping of motional mode. We also assume an additional fidelity loss per selection of $P_{\text{error}}=5\times10^{-3}$. As shown in Fig.~\ref{fig:error}(a), the expected shift of the damped apparatus is smaller than in the ideal case. While in this particular parameter regime the only observable trade-off stems from imperfect projection, the trade-off due to quantum noise (first type) can be observed at small critical $N$ by reducing the success probability of SQC. Also, we impose the same local dephasing on both the control and target qubits for the quantum search algorithm, which results in a fidelity drop due to purity loss. As a calibration, we evaluate  the algorithm under the same noise parameters in Fig.~\ref{fig:error}(b,c), confirming the presence of both types of trade-off mechanisms.

\subsection{Robustness against state selection error}

For completeness, we also consider  state selection errors, which highlight another advantage of SQC.
From an ensemble perspective, the robustness problem is equivalent to asking whether the average of weak values from an ensemble of $N$ copies of system and apparatus (with perturbation on selected state)converges to the ideal weak value.. This question was answered in the 1990s when the concept of weak measurement was introduced in Ref. \cite{pra41}. It was proved that the expected average of weak values equals the ideal weak value, with a standard deviation suppressed by $1/\sqrt{N}$. Note that this ensemble requires all $N$ post-selections to succeed, which occurs with the same overall probability as SQC.

To make this argument more concrete, we model projection errors using the language of SQC. We parameterize the pre- and post-selected states on the Bloch sphere as
	\begin{eqnarray}
	|i\rangle &=&\left[\cos{\left(\frac{\pi/2-\theta+\phi_1}{2}\right),\sin{\left(\frac{\pi/2-\theta+\phi_1}{2}\right)}}e^{i\beta_1}\right]^T,\\
		\langle f|&=&\left[\cos{\left(\frac{\pi/2-\theta+\phi_2}{2}\right),\sin{\left(\frac{\pi/2-\theta+\phi_2}{2}\right)}}e^{i\beta_2}\right],
	\end{eqnarray}
	where $\theta$ is the angle between ideal pre-selection and post-selection, and $\phi_i$  and $\beta_i$  ($i=1,2$)  quantify longitudinal and latitudinal deviations, respectively.  Assuming small errors, the weak value becomes \cite{jozsa}
	\begin{eqnarray}
		\sigma_{wx}=-\frac{2i-\left(\beta_1-\beta_2\right)+\left(\beta_1+\beta_2\right)\sin{\theta}}{\left(-2i+\beta_1-\beta_2\right)\cos{\theta}-i\left(\phi_1-\phi_2\right)\sin{\theta}},
	\end{eqnarray}
which reduces to the ideal weak value $\sec{\theta}$ as $\phi_i,\beta_i\ \rightarrow\ 0$. The physical picture is quite clear, where the longitudinal and latitudinal errors contribute to the deviation in the real and imaginary parts, respectively, of the complex weak value. Importantly, complex weak values lead to non-unitary evolution of the apparatus, represented by a Gaussian wavefunction.  As discussed in Ref. \cite{PRA76}, the apparatus undergoes the following transformation under post-selection
	\begin{eqnarray}
		\left\langle q\right\rangle_f&=&\left\langle q\right\rangle_i+gTa,\\
		\left\langle p\right\rangle_f&=&\left\langle p\right\rangle_i+2gTbVar_p,
	\end{eqnarray}
	where $a$ and $b$ are the real and imaginary part of the weak value, if the apparatus wavefunction is real-valued. The imaginary part  evolves on the apparatus as a non-unitary operation, resulting in an amplitude change (size variation) of the apparatus after post-selection. In this scenario, the deviations caused by the selection process are averaged out over $N$ rounds of SQC, provided that the magnitudes of errors follow Gaussian distributions. Consequently, the standard deviation is reduced by $1/\sqrt{N}$.

For the quantum search algorithm, projection errors can be treated analogously in a similar manner. Here, the apparatus is a qubit. A real part of the weak value introduces rotation  rotates of apparatus wavefunction around the Y-axis, while a complex weak value causes an increase or decrease in the system size,again indicating non-unitary evolution. Consequently, the expectation values of Pauli observables after post-selection evolve as
\begin{eqnarray}
	\langle \sigma_y \rangle_f&=& 	\langle \sigma_y \rangle_i+2gb [1-	\langle \sigma_y \rangle_i^2],\\
		\left\langle\sigma_x\right\rangle_f&=&\left\langle\sigma_x\right\rangle_i+2ga\left\langle\sigma_z\right\rangle_i-2gb\left\langle\sigma_x\right\rangle_i 	\langle \sigma_y \rangle_i,\\
		\left\langle\sigma_z\right\rangle_f&=&\left\langle\sigma_z\right\rangle_i+2ga\left\langle\sigma_x\right\rangle_i-2gb\left\langle\sigma_z\right\rangle_i	\langle \sigma_y \rangle_i,
	\end{eqnarray}
	demonstrating that robustness is retained in this scenario as well. Additionally, common errors such as pulse-length fluctuations,  $g\left(t\right)\rightarrow g\left(t\right)\left(1+\delta\right)$, are similarly averaged out by SQC.
In a word,   SQC mitigates both selection and systematic errors by distributing them across  $N$ rounds of selections, rather than accumulating them. This significant advantage distinguishes SQC from other protocols and underscores its promise in quantum control and computation.

\subsection{Difference between SQC and STA}

We claim that SQC resembles STA \cite{review2021} in both cases considered, as the apparatus is transported or rotated without excitation in the weak measurement regime. In effect, SQC mimics the outcome of STA by accelerating an adiabatic-like process. However, their underlying mechanisms are fundamentally different.

Both AQC and STA rely on an adiabatic reference. In AQC, the time-dependent Hamiltonian varies slowly to ensure the system remains in its instantaneous ground state. In contrast, rapid evolution without STA typically induces excitations, violating the adiabatic condition. To suppress these excitations, two principal STA techniques are employed:  (i) Invariant-based inverse engineering, where a Lewis-Riesenfeld dynamical invariant allows the design of tunable Hamiltonian parameters that guide the system without excitation, provided appropriate boundary conditions are imposed, see examples in  \cite{Erik}. (ii) Counterdiabatic driving, which adds an auxiliary term to the Hamiltonian to suppress transitions \cite{An}. While an exact counterdiabatic term generally requires full diagonalization, approximate methods using local terms or nested commutators are often used in practice \cite{review2021}. In both strategies, STA ensures continuous and unitary evolution by directly engineering the Hamiltonian.

Interestingly, spin-orbit coupling can serve as such an auxiliary term to accelerate AQC in an open-loop fashion, without relying on weak measurement. However, in this case, decoupling motional and internal states post-evolution is necessary, requiring longer operation times than SQC. SQC offers a crucial distinction: it achieves periodic decoupling of the controller (system) and target (apparatus) via post-selection, avoiding the need for prolonged evolution. This contrasts sharply with STA based on spin-orbit coupling, making SQC a more efficient and versatile scheme.
Fundamentally, SQC drives the apparatus through accumulated weak values, using interaction and post-selection rather than Hamiltonian engineering. In its ensemble version, weak-measurement-induced superoscillations arise from sequentially selecting a single system. The evolution, characterized by instantaneous and discontinuous  ``kicks" to the apparatus, is inherently non-unitary-even when the weak value is real-due to the probabilistic nature of post-selection. Nonetheless, SQC achieves high-fidelity evolution with optimal post-selection, offering clear advantages over AQC in terms of energy cost and the likelihood of superoscillations. Another significant benefit is that SQC does not require knowledge of the eigensystem or an adiabatic reference, a major departure from STA.

Beyond the two scenarios discussed here, SQC allows greater flexibility than STA in designing both the weak value and the interaction type. When the system is not coupled via momentum or angular momentum, the resulting superoscillatory function no longer behaves as a simple translation or rotation operator, thus falling outside the conventional STA framework.

\subsection{Absence of orthogonality catastrophe}

Finally, we are concerned about the selection probability 
 $|\langle i|f\rangle|$ and reckon that SQC is not applicable (even in principle) to control of large quantum systems due to the orthogonality catastrophe  \cite{Anderson}. Indeed, Anderson's orthogonality catastrophe refers to the phenomenon wherein a  local perturbation in a gas of $N$ spinless fermions, leads to the overlap between the perturbed and unperturbed many-body wavefunctions scaling as $\chi=|\langle\Psi(x_1,x_2,\cdots,x_N)|\Phi(x_1,x_2,\cdots,x_N)\rangle|\propto N^{-\alpha/2}$  with $\alpha$ characterizing the perturbation strength.

While the orthogonality catastrophe is a genuine effect in large quantum systems, it does not pose a fundamental obstacle for SQC,  though it does arise as the consequence of the quantum speed limit \cite{Mossy}.   Crucially, SQC does not require selecting on the large system itself.  As emphasized before, when applying SQC to control a large quantum system (the target), this system acts as the apparatus that accumulates the weak value.  The post-selection, which determines the probability $\langle i|f\rangle$,  is performed on the controller (the system being weakly measured), and is independent of the size of the apparatus (the large quantum system being controlled). This conceptual is made clear in Appendix \ref{MSgate}, where we consider control of high-dimensional systems. There, we construct higher-dimensional angular momentum operators as observables of the apparatus, while pre- and post-selection is still carried out on a two-level system. The resulting weak value $S_w$ induces dynamics on the large system through a sequence of weak kicks, each described by the unitary 
$\exp{\left(-igTS_w\hat{J}_y/N\right)}$.

Therefore, SQC circumvents the orthogonality catastrophe by design. It remains applicable to large-scale quantum systems, just as other control techniques, such as optimal control theory and STA do. What sets SQC apart is that it constitutes a novel framework: it drives the evolution of a target quantum system through sequential selections on another quantum system (as in the quantum search example) or on a separate degree of freedom (as in the atomic transport example).

\section{Conclusion and outlook}

In summary, we have introduced a general SQC framework and demonstrated its efficiencies and robustness with trapped ions. 
With two examples as atomic transport and quantum search, we have shown that SQC can speed up conventional adiabatic control, mimicking shortcuts-to-adiabaticity. Numerical simulations demonstrate that SQC has advantages in terms of energy cost, even when considering the probability of occurrence. We have also extended the SQC protocol to open quantum systems, where noise affects its performance, resulting in two types of trade-offs between selections and fidelity.

While our focus is on the trapped ion system, insights can be gained from the cold atoms in spin-dependent optical potentials~\cite{Martin}, and electrons in semiconductor quantum dots~\cite{Sherman} with scaling approach.  In principle, SQC can be applied to control physical systems with interactions, encompassing not only spin-orbit coupling but also spin-spin interactions and other relevance. Additionaly, our work can be extended to coherent matter-wave splitting, as we have presented the analytical formulation of the apparatus after post-selection. Furthermore, SQC for collective behaviour has applications in the emergence or suppression of superradiant phase transition in the Dicke model~\cite{dicke}. In these cases, system and apparatus are no longer coupled to construct translation or rotation operator, i.e., do not correspond to accelerated adiabatic processes anymore. We believe that SQC can provide a deeper understanding of quantum foundations and offer potential for various quantum technologies.

\section*{acknowledgment}
This work is supported by the Basque Governmentthrough Grants No. IT1470-22, the project grants PID2021-126273NBI00, PID2021-123131NA-I00 funded by MCIN/AEI/10.13039/501100011033, by ``ERDF a way of making Europe", and by the European Union Next Generation EU/PRTR".  Y. D. thanks the European Commission for a Marie Curie PF grant (No. 101204580 FELQO).

\appendix

\section{Weak-to-strong quantum measurement, weak value, and expectation value}
\label{Weakvaluecriteria}

\begin{figure}
	\includegraphics[width=0.5\textwidth,height=0.42\linewidth]{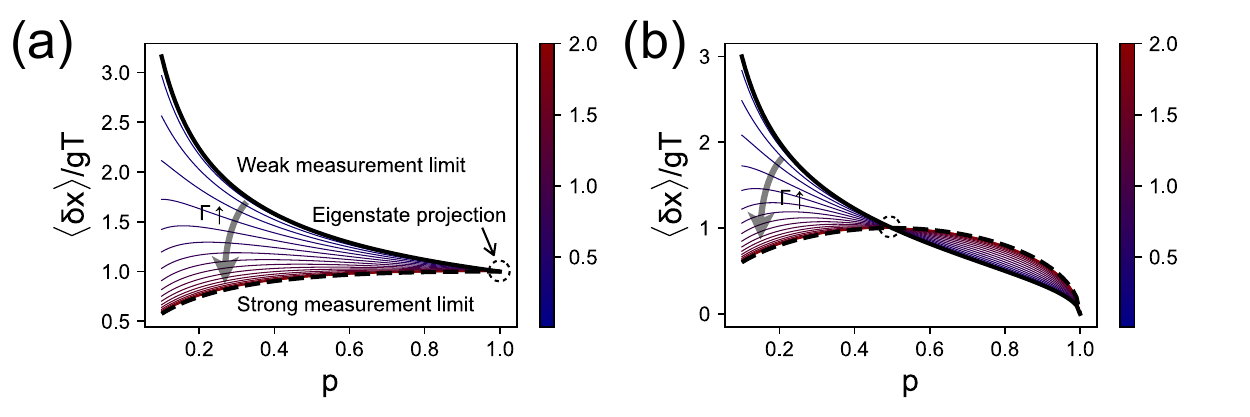}
	\caption{\label{fig:w2s} The relative central position shift from the initial Gaussian state to the cat state $\langle\delta x\rangle/gT$ as a function of the selection probability $p$ with different measurement strength characterized by the interference factor $\Gamma=gt/(\sqrt{2}x_0)$. (a) We set the states for pre-selection and post-selection  (\ref{eqn:if}). The thick black solid curve denotes the weak value readout $\sigma_{xw}=1/\sqrt{p}$ in the weak measurement limit $\Gamma\rightarrow0$. The thick black dashed curve denotes the bizzare readout $2\sqrt{p}/(1+p)$ in the strong measurement limit $\Gamma\rightarrow\infty$, which is not in accordance with the expectation value $\langle f|\hat{\sigma_x}|f\rangle=\sqrt{p}$. Other possible readout curves lie between them in the color according to the value of $\Gamma$. We circle the point for eignenstate projection of $|+\rangle$, where the weak and strong measurement meet each other losslessly. (b) We set the states for pre-selection and post-selection be $|i\rangle=|0\rangle$ and $|f\rangle=\sqrt{p}|0\rangle+\sqrt{1-p}|1\rangle$, respectively, for equalizing the expectation value and the readout in the strong measurement limit as $2\sqrt{-(-1+p)p}$. The weak measurement output is $\sqrt{1-p}/\sqrt{p}$. Here the eigenstate projection has a probability of $0.5$. Other configurations definitions are the same as those in the previous figures.}
\end{figure}

In this Appendix, We now present a detailed calculation of the weak-to-strong quantum measurement transition to verify the necessity of using SQC instead of relying on a single-shot strong measurement.  To begin, we consider pre-selected  and post-selected states, expressed  in the eigenbasis of the operator, $\hat{\sigma_x}$, for measurement, e.g.,
$|i\rangle=a_0|+\rangle+a_1|-\rangle$ and $|f\rangle=\tilde{a}_0|+\rangle+\tilde{a}_1|-\rangle$. 
We model the quantum measurement process for arbitrary measurement strength by coupling to a Gaussian pointer state $|\Psi(x)\rangle$,  such that the joint system evolves as
\begin{equation}
	\label{eqn:separate}
	\langle f|e^{-igT\hat{\sigma}_x\otimes\hat{p}}|i\rangle \otimes |\Psi(x)\rangle = c_0|\Psi(x{-}gT)\rangle + c_1|\Psi(x{+}gT)\rangle,
\end{equation}
where $c_0=a_0\tilde{a}_0^*$ and $c_1=a_1\tilde{a}_1^*$. These two spatially displaced wave packets can be renormalized to form a quantum superposition (a Schr\"odinger cat state):
\begin{equation}
	|\text{cat}\rangle = \frac{c_0|\Psi(x{-}gT)\rangle + c_1|\Psi(x{+}gT)\rangle}{\sqrt{|c_0|^2 + |c_1|^2 + 2\mathrm{Re}(c_0^* c_1 \langle\Psi(x{-}gT)|\Psi(x{+}gT)\rangle)}}.
\end{equation}
To quantify the quantum interference between the two displaced wave packets, we use the overlap $\langle\Psi(x-gT)|\Psi(x+gT)\rangle=\exp(-\Gamma^2)$, where the interference factor is defined as $\Gamma=gT/(\sqrt{2}x_0)$, with $x_0$ the standard deviation of the initial Gaussian. The measurement outcome is given by the shift in the expectation value of the pointer position:
\begin{eqnarray}
	\label{eqn:cat}
	\delta x=\langle\text{cat}|\hat{x}|\text{cat}\rangle = \frac{(|c_0|^2-|c_1|^2)gt}{|c_0|^2+|c_1|^2+(c_0^*c_1+c_0c_1^*)e^{-\Gamma^2}}.
\end{eqnarray}
In the weak-coupling limit  ($\Gamma \rightarrow 0$), this shift reduces to 
\begin{eqnarray}
	\langle\delta x\rangle|_{\Gamma\rightarrow0}=\frac{(|c_0|^2-|c_1|^2)gt}{|c_0|^2+|c_1|^2+c_0^*c_1+c_0c_1^*}=\sigma_{xw}gt,
\end{eqnarray} 
which is precisely the weak value of $\hat{\sigma_x}$. 
In contrast,  in the strong-coupling regime  ($\Gamma \rightarrow \infty$),  the interference vanishes and the readout becomes $\langle\delta x\rangle|_{\Gamma\rightarrow\infty}=(|c_0|^2-|c_1|^2)gt/(|c_0|^2+|c_1|^2)$ does not always amplifies the expectation value $\langle f|\hat{\sigma_x}|f\rangle$. For instance, in the main text, we chose the pre- and post-selection states (see Eq.~\eqref{eqn:if}) to realize a weak value $\sigma_{xw}=1/\sqrt{p}$ while the expectation value becomes 
 $\langle f|\hat{\sigma_x}|f\rangle=\sqrt{p}$. In the strong measurement limit, however, the readout instead yields $2\sqrt{p}/(1+p)$, which deviates from both values. To illustrate this, we artificially construct a selection:
\begin{eqnarray}
	|i\rangle=|0\rangle, |f\rangle=\sqrt{p}|0\rangle+\sqrt{1-p}|1\rangle,
\end{eqnarray}
which gives a weak value $\sigma_{xw}=\sqrt{1-p}/\sqrt{p}$ and an expectation value 
$2\sqrt{-(-1+p)p}$, now in agreement with the strong measurement readout. We show the weak-to-strong transition under both settings in Fig.~\ref{fig:w2s}. This demonstrates how weak and strong measurement regimes generally yield different outcomes, and why SQC within the weak measurement regime is essential for reliable and controllable evolution. 

	\begin{figure}
	\includegraphics[width=8.6cm]{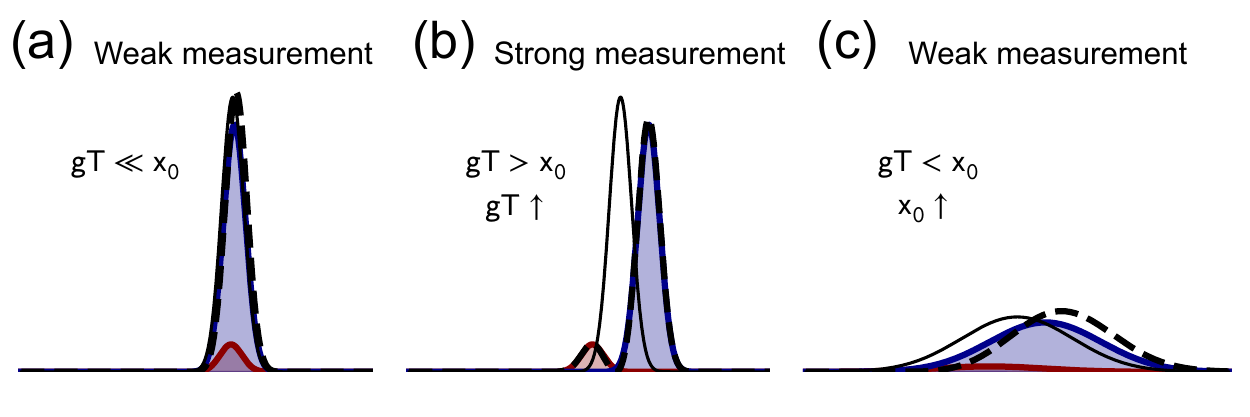}
	\caption{\label{fig:regimes}Regimes of quantum measurement. The Gaussian apparatuses are plotted by thin black curve, resulting in two sub-wave packets $|\Psi(x+gT)\rangle$ (red) and $|\Psi(x-gT)\rangle$ (blue) after coupling and post-selection. With quantum interference, one has the $|\text{cat}\rangle$ state plotted by thick black dashed curves. We plot the absolute squares of all wave functions. Dimensionless parameters: (a) $gT=0.25,~x_0=1.58$; (b) $gT=4,~x_0=1.58$; (c) $gT=4,~x_0=7.91$}
\end{figure}

Next, we analyze the weak-to-strong transition and optimal design of selections. Here, we depict the difference between weak measurement and strong measurement, which is described by Eq.~\eqref{eqn:separate} and~\eqref{eqn:cat}. Fig.~\ref{fig:regimes}(a) demonstrates the weak measurement in the most well-known way. As the main text said, one should have an inaccurate Gaussian apparatus that is spatially wide, which is characterized by its standard deviation. In this regime, the criteria for weak measurement $gT\ll x_0$ is satisfied. Two sub-wave packets in Eq.~\eqref{eqn:separate} are overlapped, transporting the apparatus slightly with quantum interference. Once the coupling strength $gT$ becomes larger, they no longer overlap but separate, showing a strong measurement with $gT>x_0$ in Fig.~\ref{fig:regimes}(b). Although the coupling strength is large, one can still realize a weak measurement by having a more inaccurate Gaussian apparatus, as given in Fig.~\ref{fig:regimes}(c). The spatially wider sub-wave packets overlap again, showing the transplant-like behaviour with quantum interference again.

\section{ Analog simulation of quantum search with M\o lmer-S\o rensen gate and its generalization}

\label{MSgate}

In the Appendix, we propose the use of a YY interaction for recovering Grover's search through sequential selections. As mentioned earlier, this interaction can be simulated analogously using two trapped ions. By bichromatically exciting the ions, we obtain the time-dependent Hamiltonian $
	H_\text{bic}(t)=\frac{\hbar}{2}\Omega\left(\hat{\sigma}_+^{(1)}+\hat{\sigma}_+^{(2)}\right)\left[1+i\eta\left(\hat{a}e^{-i\nu t}+\hat{a}^\dagger e^{i\nu t}\right)\right]\left(e^{-i\delta t}+e^{i\delta t}\right)+\text{H.c.}$,
which yields the exact propagator:
\begin{equation}
	U_\text{bic}(t)=\hat{D}\left\{\frac{\eta\Omega}{\epsilon}\left(e^{i\epsilon t}-1\right)\hat{J_y}\right\}e^{i[\frac{\eta^2\Omega^2}{\epsilon}t-\frac{\eta^2\Omega^2}{\epsilon^2}\sin\left(\epsilon t\right)]\hat{J_y^2}}.
\end{equation}
Here, $\epsilon=\nu-\delta$ is the detuning from the motional sidebands,  $\hat{D}(\alpha)$  denotes the displacement operator, and the collective spin operator is defined as 
 $\hat{J_y}=\hat{s_y}^{(1)}+\hat{s_y}^{(2)}$. When the interaction time is set to  $t_\text{MS}=2\pi/\epsilon$,
the displacement operator vanishes, and the evolution corresponds to the action of an effective Hamiltonian:
\begin{equation}
H=-\hbar(\eta^2\Omega^2/\epsilon)\hat{s_y}^{(1)}\otimes\hat{s_y}^{(2)},
\end{equation}
which entangles the controller qubit $\hat{s_y}^{(1)}$ with  the target qubit $\hat{s_y}^{(2)}$.  For different initial database states, one can always identify a suitable rotation axis, requiring interactions of the form $\hat{\sigma}_\phi^{1}\otimes\hat{\sigma}_\phi^{2}$, where $\hat{\sigma}_\phi=\cos\phi\hat{\sigma_x}+\sin\phi\hat{\sigma_y}$. Such generalized couplings are feasible by adjusting the phase of the bichromatic laser fields.
Note that this analog simulation does not satisfy the energetic bound discussed in the main text, as it does not implement a direct YY interaction but rather realizes it effectively through second-order dynamics involving the motional degrees of freedom.

For entries with $N_G>2$,   the database requires multiple qubits to encode the quantum information.Consequently, the simple Bloch sphere picture used in the minimal two-state model is no longer valid, necessitating a generalization of the rotation scheme to a higher-dimensional Hilbert space.
Assume that the initial state for the quantum search  of target state $|t\rangle$ is given by
\begin{equation}
	|\Psi\rangle=\sin\frac{\theta}{2}|t\rangle+\cos\frac{\theta}{2}|\tilde{t}\rangle,
\end{equation}
where $|\tilde{t}\rangle=\sum_{N_G-1}c_i|i\rangle$  is the normalized superposition of all non-target states, orthogonal to $|t\rangle$. To implement the SQC protocol for quantum search in this larger space, we need to construct a generalized angular momentum operator $\hat{J}$ such that $\exp(-i\phi\hat{J})|\Psi(\theta/2)\rangle=|\Psi(\theta/2+\phi)\rangle$, i.e., $\hat{J}$ generates a rotation in the two-dimensional subspace spanned by $|t\rangle$ and $|\tilde{t}\rangle$, analogously to the rotation in the minimal model. If all basis amplitudes are real, we can define an effective Y-axis of rotation. 
Specifically, we introduce effective ``plus" state and ``minus" state
\begin{eqnarray}
	|\tilde{\pm}\rangle=\frac{\sqrt{2}}{2}|t\rangle \pm \frac{\sqrt{2}}{2}|\tilde{t}\rangle,
\end{eqnarray}
 from which we construct the effective angular momentum operator:
\begin{equation}
	\hat{J_y} =-\frac{i}{2}(|\tilde{-}\rangle\langle\tilde{+}|-|\tilde{+}\rangle\langle\tilde{-}|),
\end{equation}
which rotates the parameterized wave function as we desire.

\end{document}